\documentstyle[prb,aps,epsfig]{revtex}

\newcommand{\be}{\begin{equation}}
\newcommand{\ee}{\end{equation}}
\newcommand{\bea}{\begin{eqnarray}}
\newcommand{\eea}{\end{eqnarray}}

\newcommand{\p}{\partial}

\newcommand{\rd}{\mbox{d}}
\newcommand{\ri}{\mbox{i}}

\newcommand{\nn}{\nonumber\\}

\begin{document}

\draft
\title{Superconductivity and Charge Density Wave in a Quasi-One-Dimensional Spin Gap System}

\author{Sam T. Carr$^{1,2}$ 
and  Alexei M. Tsvelik$^1$}

\address{$^1$Department of  Physics, Brookhaven National Laboratory,
 Upton NY 11973, USA \\
 $^2$Department of Theoretical Physics, Oxford University, 1 Keble Road, Oxford, UK}

\date{\today}

\maketitle

\begin{abstract}
We consider a model of  spin-gapped chains  weakly coupled 
by  Josephson and Coulomb interactions.  Combining such non-perturbative methods as  bosonization and Bethe ansatz to treat the intra-chain interactions 
 with  the Random Phase Approximation for the inter-chain couplings and
the first corrections to this, we
investigate the phase diagram of this model. The phase diagram shows both charge
density wave ordering and superconductivity.  These 
phases are separated by line of critical points which exhibits an approximate an SU(2) symmetry. We consider the effects of a magnetic field on the system.
We apply the theory to the material
${\rm Sr_2 Ca_{12} Cu_{24} O_{41}}$ and suggest further experiments.

\end{abstract}

PACS numbers: 71.10.Hf, 74.20.Mn 

\sloppy

\section{Introduction}

 Quasi-one-dimensional (1D) models  are often used to test various 
theoretical ideas in the area of strongly correlated electron systems for
 the simple
 reason that most known non-perturbative approaches work only in one dimension.
\cite{boz,eme79} The route often taken is to use a non-perturbative
 solution of a strictly one-dimensional model and then use mean field or
 the Random Phase Approximation (RPA) to take into account the inter-chain
 interactions.  Through techniques such as bosonization and Bethe ansatz,
many results are known about such  one-dimensional systems as spin chains and Tomonaga-Luttinger liquids  which form the skeleton of these quasi-one-dimensional models.
Linking these using the RPA formalism has yielded many successful experimental
predictions, for example for linear conductors\cite{fb81} and 
 for magnetic systems.\cite{sch96,etd97,betg01}  A very early use of this
technique is Efetov and Larkin\cite{el74,el75} who estimated the transition
temperatures in the same model we use.  

 As it is well known, the RPA formally represents  the leading term   in a
 perturbation expansion in $1/z_\perp$,  where $z_\perp$ is
the number of nearest-neighbour chains in the lattice.
 For real experimental systems this number is not usually  large so it is important to know 
 about higher order  contributions in $1/z_\perp$.  The recent results for
 the quasi-one-dimensional Heisenberg magnets indicate that the
 worst these corrections can do is about a $25\%$ shift in the
 transition temperature.
\cite{ik00,boc01} The relative smallness of these corrections
 demonstrates the validity
of the RPA approximation when considering real systems; in our case it turns
out that the corrections are even smaller.

In this paper we follow the same road  
and discuss a simple model of a non-BCS superconductor. In the model we consider the formation of superconducting pairs  on one-dimensional chains is triggered by formation of  a spin gap. The three-dimensional coherence is established through  the inter-chain Josephson coupling. Since the latter  coupling competes with the Coulomb interaction, which can destroy the superconductivity and establish Charge Density Wave (CDW) ordering. As we shall show,  these two
 phases are separated by a critical line with increased symmetry. Near this
line, we take into account the interplay between these two interactions 
 considering corrections to RPA.

 The  model we use has been considered in some detail recently
\cite{coke00} in the context of high-$T_c$
 superconductivity. It was assumed that the one-dimensional behaviour came about from the
formation of stripes.\cite{ekt99}  Since  in the stripe picture, fluctuations
of the stripes dephase the CDW coupling,\cite{ekz99}  only the SC inter-chain
interaction was considered. In our paper we retain the Coulomb interaction and  therefore expect it to be relevant  to materials
that are structurally quasi-one-dimensional such as the Bechgaard salts or
some cuprate materials such as the family ${\rm Sr_{14-x} Ca_{x} Cu_{24} O_{41}}$.

In section \ref{model} we introduce the model we will be dealing with.  In
section \ref{effsu2} we show that this model has an SU(2) symmetric 
quantum critical line.  In section \ref{RPA1} we 
calculate the transition temperature for the model within the RPA
 approximation.  Treating the inter-chain coupling in the mean field
 approximation we obtain an effective sine-Gordon model for each chain.
 Using the exact results for this model we calculate the zero temperature spectral gap $M$ and derive the expression for the ratio $T_c/M$.  Here, we also consider the properties of our system in a magnetic field.  In section 
\ref{RPA2} we look at the first corrections to RPA which gives us an improved
phase diagram of the model.  In section \ref{rpa2d} we show that
the same general behaviour also occurs in two dimensions, although the
transition here is Kosterlitz-Thouless rather than the symmetry breaking
found in higher dimensions.  Finally, in section \ref{expt}, we show that the
quasi-1D compound ${\rm Sr_2 Ca_{12} Cu_{24} O_{41}}$ is a beautiful example
of our model and we discuss the measured properties of it in relation to
our theory.  We also make some quantitative predictions about this material
which could be confirmed by further experiments.

\section{The model}\label{model}

 Let us  consider a system of conducting one-dimensional units (we will call them {\it chains}, though in reality they may be, for instance, ladders) weakly  coupled to each other.  As often happens in one dimension, the spin and charge
degrees of freedom decouple at low energies.  We assume that the spin
sector aquires a gap, and that the filling of each individual
 chain is incommensurate with the 
lattice so that the low energy behaviour of the charge sector is decribed universally by the Gaussian model.  The Hamiltonian density is therefore
\bea
&&{\cal H}_{\rm chain}= {\cal H}_{\rm charge} + {\cal H}_{\rm spin},\\
&&{\cal H}_{\rm charge} = \frac{1}{2}[K_c(\p_x\Theta)^2 + K_c^{-1}(\p_x\Phi)^2]
\eea
where $[\Theta(x),\Phi(y)] = \ri\theta(x - y)$.  We don't write down the form
of the Hamiltonian in the spin sector: our only requirement is that it has a 
gap $\Delta_s$.
\footnote{In a specific case of single chains a realistic description
 of the spin sector is given by the SU(2) Thirring model
 Hamiltonian: \cite{boz}
\bea
{\cal H}_{\rm spin} = \frac{2\pi v_s}{3}(:J^aJ^a: + :\bar J^a\bar J^a:) -  g :J^a\bar J^a:\label{spin}
\eea
where  $v_s$ is the spin velocity and 
$J^a, \bar J^a$ are chiral SU(2) currents satisfying the level $k = 1$ SU(2) Kac-Moody algebra. The spin gap is generated when  $g > 0$ such that the current-current interaction in the spin sector is marginally {\it relevant}. In the case of ladders a description of the spin sector is more complicated; this however does not affect the charge Hamiltonian and therefore will not concern us here.}

% In the  Lagrangian description it is convenient to bosonize  the spin sector;
%this results in  the sine-Gordon Lagrangian with the density
%\bea
%&&{\cal L}_{\rm charge} = \frac{1}{2K_c}(\p_{\mu}\Phi_c)^2,\\
%&&{\cal L}_{\rm spin} = \frac{1}{2}(\p_{\mu}\Phi_s)^2 + 
%\Lambda \cos(\sqrt{8\pi K_s}\Phi_s), ~~~ K_s = 1 - g/2\pi.
%\eea

% The SU(2) Thirring model is well understood; the exact results exist for its 
%thermodynamics and correlation functions. The current-current interaction open%s a spectral gap $\Delta_s$ in the spin sector. 

The spin  gap blocks single-particle tunneling processes
 between the chains. Then the multi-particle processes generate pair hopping. In what follows we shall 
assume that the inter-chain tunneling matrix element is much smaller than the 
spin gap. In this case one can take into account only two-particle virtual
 processes giving  rise to Josephson coupling between the chains. They lead to the following Hamiltonian  
\bea
{\cal H}_{\rm sc} = \frac{1}{2}J_{\rm eff}\sum_{n \neq m}
 :\cos[\sqrt{2\pi}(\Theta_n - \Theta_m) - 2e H b_{nm}x/c]: \label{scint}
\eea
where the dots signify that operators are normal ordered with respect to the state with spin gap and therefore the ultraviolet cut-off for the correlation functions of bosonic exponents is $\Delta_s$. The fields without index are assumed to be from the charge sector, as will be
the case from here on. We have also introduced external magnetic field $H$ directed perpendicular to the chains; $b_{nm}$ is the projection of the inter-chain  lattice vector on the direction perpendicular both to the chains and the magnetic field. 

 An analysis of dimensionalities as shown in Appendix \ref{vjest} yields 
\be
J_{\rm eff} \sim \left(\frac{\Delta_s}{\Lambda}\right)^{1/K_c-1}
\frac{t^2}{\Delta_s}
\ee
where $t$ is the single particle hopping and $\Lambda$ is related to the original bandwidth. 

 Interaction (\ref{scint}) has scaling dimension 
\bea
d_{\rm sc} = 1/(2K_c)
\eea
and therefore is relevant even for repulsive interactions in the charge sector
 provided they are not too strong ($K_c > 1/2$). This is a well known effect
 of the spin gap; it generates preformed pairs making it easy for them to
 condense. \cite{ek95}

There is also a Coulomb interaction between the two chains
In the spin gap regime, there is only one term in here that remains
relevant: it is the coupling of $2k_F$-components of the charge
 density which gives the effective Hamiltonian density (Appendix \ref{vjest})
\bea
{\cal H}_{\rm cdw} = \frac{1}{2}V_{\rm eff}\sum_{n \neq m}
 :\cos[\sqrt{2\pi}(\Phi_n - \Phi_m)]: 
\eea
where
\be
V_{\rm eff} \sim \left(\frac{\Delta_s}{\Lambda}\right)^{K_c} V_0.
\ee
The corresponding scaling dimension is 
\bea
d_{\rm cdw} = K_c/2.
\eea

 The effective action for coupled chains is therefore 
\bea
{\cal L}_{\rm eff} = \frac{1}{2K_c}\sum_n (\p_{\mu}\Phi_n)^2 +
 \frac{1}{2}\sum_{n \neq m}
 \{ V_{nm}:\cos[\sqrt{2\pi}(\Phi_n - \Phi_m)]: +
 J_{nm}:\cos[\sqrt{2\pi}(\Theta_n - \Theta_m - 2e H b_{nm}x/c)]:\} \label{inter}
\eea
and has $\Delta_s$ as the ultraviolet cut-off.  We will be considering
nearest-chain interactions only, i.e. $V_{nm}=V, J_{nm}=J$ for
 neighbouring chains and zero otherwise.
In what follows we will be most interested in the case
 $K_c \approx 1$ when both interactions are important.

\section{An Effective theory of the Critical Point}\label{effsu2}

For a general value of $K_c$ the symmetry of the model is U(1)$\times$U(1) which corresponds to 
independent global shifts of $\Phi$ and $\Theta$. When   $K_c =1$ and $V = \pm J$ the symmetry increases and becomes SU(2).
 To see this we use the non-Abelian bosonization description.\cite{boz,wit84}
 At $K_c = 1$ the exponents $\exp[\pm \ri\sqrt{2\pi}\Phi],\exp[\pm \ri\sqrt{2\pi}\Theta]$ have conformal dimensions (1/4,1/4) and can be understood
 as matrix elements of the tensor field $g_{ab}$ from the S=1/2 representation
 - the first primary field of the   level $k =1$ Wess-Zumino-Novikov-Witten
 model (for a discussion of this model, see e.g. Itzykson and Drouffe\cite{id89}):
\bea
\hat g = \left(
\begin{array}{cc}
\exp[\ri\sqrt{2\pi}\Phi] & \exp[\ri\sqrt{2\pi}\Theta]\\
\exp[-\ri\sqrt{2\pi}\Theta] & \exp[-\ri\sqrt{2\pi}\Phi]
\end{array}
\right).
\eea

The Gaussian part of the action becomes the sum of the WZNW actions from individual chains:
\bea
\frac{1}{2}\sum_n (\p_{\mu}\Phi_n)^2 \rightarrow \sum_n W[g_n].
\eea
and the interaction term in (\ref{inter})  can be written as 
\bea
L_{int} = \sum_{n \neq m}\{(V - J)\sum_{a = 1,2} [g_n^{(aa)}[g_m^+]^{(aa)} +(n \rightarrow m)] + J\mbox{Tr}(g_ng_m^+ + g_mg_n^+)\}. \label{coupling} 
\eea
This description is convenient since it contains only mutually local fields and therefore can be considered as the Ginzburg-Landau theory. 

In three spatial dimensions the system undergoes a phase transition into 
 the ordered state where the matrix $g$ acquires an average value throughout the system. In the long wave limit  one can replace the last term in (\ref{coupling}) by 
\bea
(\p_y g)(\p_y g^+)
\eea
and omitting the time dependence of the fields we 
obtain the following  Ginzburg-Landau free energy:
\bea
F = b^{-2}\int \rd x\rd^2 r \mbox{Tr}[\frac{va_0}{16\pi}(\p_x g^+\p_x g) + Jb^2(\nabla_{\perp}g^+\nabla_{\perp}g)] + F_{anisotropy}
\eea
where $b$ is the lattice constant in the transverse direction and 
\bea
 F_{anisotropy} = (V -J)b^{-2}\int \rd x\rd^2 r \sum_{a = 1,2} g^{(aa)}[g^+]^{(aa)}. 
\eea

We can now re-parameterize the theory.
 The order parameter is the SU(2) matrix $g$. Its relation to the CDW and SC phases 
$\Theta$ and $\Phi$ are:
\bea
g = \exp[\ri \sigma^3(\Phi + \Theta)/4]\exp[\ri \sigma^1\alpha/2]
\exp[\ri \sigma^3(\Phi - \Theta)/4].
\eea
The Ginzburg-Landau free energy density is
\bea
 {\cal F} = \frac{1}{2}\rho[ \cos^2(\alpha/2)(\nabla\Theta)^2 + \sin^2(\alpha/2)(\nabla\Phi)^2 
] +  \frac{1}{2}\rho(\nabla\alpha)^2 + (V - J)\cos\alpha.\label{glvj} 
\eea
This is interpreted as follows: when $V-J$ is positive, $\alpha$ is pinned at
$\pi$ so that the coefficient in front of $(\nabla\Phi)^2$ is non-zero and
hence $\Phi$, the CDW order parameter, is constant throughout the material.
When $V-J$ is negative, $\alpha$ is pinned at $0$ and hence it is $\Theta$,
the superconducting order parameter that acquires an expectation value.
When $V-J=0$ we are at the critical point where the free energy of the
superconducting and insulating phases becomes equal.  The effects of this
 $V-J$ mode will be considered throughout the rest of the paper.

\section{Phase diagram in magnetic field and critical temperature}\label{RPA1}

 For two chains the problem was solved by
 Shelton {\it et al.}\cite{snt95}  There are two modes; one symmetric in
the two chains and the other antisymmetric. 
In the presence of the inter-chain interactions, the symmetric mode remains 
 gapless and
 the antisymmetric sector splits into two Majorana fermions with gaps (V + J)
 and (V - J).

For an infinite number of chains, we expect to see a similar sort of behaviour.
The gapless symmetric mode in the the case of two chains will in some sense
be the Goldstone mode in our infinite system and we expect to see a range
of other modes with gaps ranging from $V-J$ to $V+J$.  We will see that within
the basic RPA approximation we cannot reproduce this behaviour: the properties
will depend on the stronger of $V$ and $J$ but not both.  However when we go
beyond the first order term we can start probing the interplay between these
two competing interactions.

To begin with, we estimate the critical temperature using RPA.
 Within this approximation the pairing and the CDW susceptibilities
 are given by 
\bea
\chi_{sc} = \frac{\chi^{(0)}_{sc}}{1-Jz_\perp\chi^{(0)}_{sc}},\nonumber\\
\chi_{cdw} = \frac{\chi^{(0)}_{cdw}}{1-Vz_\perp\chi^{(0)}_{cdw}} \label{susc}
\eea
where $z_\perp$ is the number of nearest neighbour chains.
These are shown diagrammatically in figure \ref{rpa} (a) and (b).

 When $K_c=1$ the bare
 susceptibilities are equal to each other  and therefore the instability
 occurs in that  channel where the interaction is stronger.  This is shown
 explicitly in Appendix \ref{mcmf}.  If $K_c\ne 1$, the instability still
occurs in the stronger channel, although this now depends not only on the
values of $V$ and $J$ but also on $K_c$ and $\Delta_s$, the crossover point being
\be
\left(\frac{t^2}{\Delta_s}v_c\right)^\frac{1}{2 - 1/2K_c}\sim \left(\frac{V}{v_c}
\right)^\frac{1}{2 - K_c/2}.
\ee

 An important modification occurs in magnetic field which affects the inter-chain interaction in the superconducting channel (\ref{scint}). In this case the susceptibilities corresponding to the  lattice directions {\bf l} should 
be taken at wave vector $2e({\bf H[\hat x \times l]})/c$, where $\hat x$ is the unit vector along the chains. Therefore the RPA criterion for the transition is replaced by 
\bea
1= \sum_l J_l\chi^{(0)}_{sc}\{q = 2e({\bf H[\hat x \times l]})/c\} \label{field}
\eea

For definiteness  let us assume that the instability occurs in the
superconducting channel
which is the most likely case for $K_c>1$.  Note that the duality
property of the effective Lagrangian (\ref{inter}) under
$K \rightarrow 1/K,~ V \leftrightarrow J,~ \Theta\leftrightarrow\Phi$ means
 that all of the results in this and the next section are identical for the
CDW channel.

 In a Tomonaga-Luttinger liquid  with the ultraviolet cut-off $\Delta_s$ the static susceptibility for the operator
 with scaling dimension $d$ is given by \cite{sb83}
\bea
\chi^{(0)}(q) = \frac{2}{\Delta_s^2} \sin \pi d \left(
\frac{2\pi T}{\Delta_s}\right)^{-2+2d} \Gamma^2(1-d)\left|\frac{\Gamma(d/2 + ivq/4\pi T)}{\Gamma(1 - d/2 + ivq/4\pi T)}\right|^2. \label{susco}
\eea
where $v$ is the velocity in the charge sector. 

\subsection{Zero magnetic field; the critical temperature and the vortex energy}
 Substituting (\ref{susco}) with $q = 0$ into Eq.(\ref{susc}) we obtain 
\bea
T_c = \frac{\Delta_s}{2\pi}
\left(
\frac{2Jz_\perp}{\Delta_s} \sin \pi d \frac{\Gamma^2 (d/2) \Gamma^2 (1-d)}
{\Gamma^2 (1 - d/2) } \right)^{\frac{1}{2-2d}}.\label{tcrpa}
\eea
The scaling properties of this equation were calculated for the first time
in reference \onlinecite{el75}.

 Below the transition temperature the long-wavelength fluctuations of superconducting order parameter are three-dimensional. The amplitude fluctuations are, however, mostly one-dimensional and their spectral weight is concentrated above certain energy which plays a role of a pseudo-gap. The zero temperature value of the pseudo-gap can be found from the  mean-field theory combined with the exact results for the sine-Gordon model. In this approach one approximates 
the inter-chain interaction
\bea
J\sum_{<nm>}\cos \beta (\phi_n - \phi_m)
\eea
( $\beta\phi = \sqrt{2\pi}\Theta$ and $\beta^2 = 2\pi K_c^{-1}$) by
\bea
 2\mu \cos\beta\phi
\eea
where
\bea
 2\mu = Jz_\perp \Delta_s \langle \cos\beta\phi \rangle.
\eea
This expectation value is known
exactly \cite{lz96}:
\bea
\langle \cos\beta\phi  \rangle
= \frac{(1+\xi) \pi \Gamma(1-d/2)}{16 \sin \pi\xi\ \Gamma(d/2)}
\left(\frac{\Gamma(\frac{1}{2}+\frac{\xi}{2})\Gamma(1-\frac{\xi}{2})}
{4\sqrt{\pi}}\right)^{(d-2)} \left( 2\sin \frac{\pi\xi}{2} \right)^d 
\left(\frac{M}{\Delta_s}\right)^d
\eea
where $M$ is the soliton mass in the SG model, and is related to $\mu$ by
\bea
\mu = \frac{\Gamma(d/2)}{\pi\Gamma(1-d/2)} \left(
\frac{2\Gamma(\xi/2)}{\sqrt{\pi}\Gamma(\frac{1}{2}+\frac{\xi}{2})}
\right)^{d-2} \left(\frac{M}{\Delta_s}\right)^{2-d} \Delta_s^2.
\eea
In all these equations, $d=\beta^2/4\pi$ is the scaling dimension of the
field $e^{i\beta\phi}$, and $\xi=1/(2-d)$.
These mean-field relations are solved to give
\bea
M = \Delta_s\left[\frac{Jz_\perp}{\Delta_s}\frac{1}{2(d-2)}
\tan\frac{\pi\xi}{2}\right]^{\frac{1}{2-2d}}
\left[ \frac{\pi\Gamma(1-d/2)}{\Gamma(d/2)} \left(
\frac{\Gamma(\frac{1}{2}+\frac{\xi}{2})\sqrt{\pi}}{2\Gamma(\xi/2)}
\right)^{(d-2)}
\right]^{\frac{1}{1-d}}.
\eea

\begin{figure}
\begin{center}
\epsfig{file=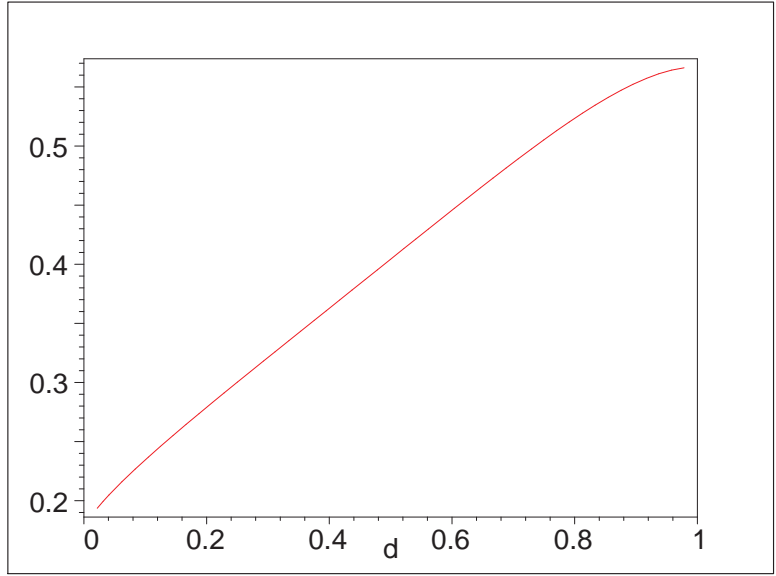,totalheight=3.0in,width=4.0in}
\end{center}
\caption{A graph of $T_c/M$ against $d$. The value  $d=1$ corresponds to the BCS limit,
decreasing $d$ corresponds to increasing repulsion.}\label{Tcvd}
\end{figure}

The ratio $T_c/M$ which is often considered in the theory of superconductivity
is plotted as a function of $d$ in figure \ref{Tcvd}.  It's numerical value
 in certain limits is:
\bea
\frac{T_c}{M}(d=0) & = & \frac{\sqrt{2}}{8} \approx 0.177, \\
\frac{T_c}{M}(d=1/2) & = & \frac{3}{16} \frac{\sqrt{3\pi}
(\Gamma(2/3)\Gamma(5/6))^3}{\Gamma(3/4)^8} \approx 0.404.
\eea
In the limit $d\rightarrow 1$
which corresponds to weak coupling, our expressions for $T_c$ and $M$ diverge
in this approximation.
However their ratio can still be evaluated.  Writing $x=1-d$ and expanding
 all the gamma functions as Taylor series in $x$ gives us the BCS value
\bea
\frac{T_c}{M}(d\rightarrow 1) = \frac{1}{2\pi} \lim_{x\rightarrow 0}
  \left[ 1 + (\ln 2 + \gamma)x\right]^\frac{1}{x}
= \frac{1}{\pi} e^\gamma \approx 0.567
\eea
where $\gamma \approx 0.57722$ is Euler's constant.  

Notice that in comparing to experiments, one has to remember that $M$ is not the single particle gap.   Single particle
 spectroscopies such as basic tunneling would see a gap much closer to
$\Delta_s$, the spin gap.  To probe $M$, one would have to look at experiments
involving pairs of electrons, such as Andreev tunneling. In the context of sine-Gordon model, $M$ is the soliton mass. Solitons correspond to spatial changes in the superconducting phase $\Theta$ and hence to vortices. Therefore $M$ is the minimal energy necessary to create a vortex.  It should also be noticed that at $d < 1$ the sine-Gordon model has not only solitons, but  bound states which, being neutral,  should be interpreted as vortex-antivortex pairs. At $d < 1/2$ the energy of the first bound state is smaller than the soliton.
See Carlson {\it et al.}\cite{coke00} for
a nice discussion of the implications of having these two energy scales.

\subsection{Phase diagram in a magnetic field}

 To keep the calculations as simple as possible, let us consider the simplest possible situation when a given chain has four nearest neighbours with Josephson couplings $J_z$ and $J_y$ and the magnetic field lays in the $yz$ plane. Combining  Eq.(\ref{field}) and Eq.(\ref{susco}) we obtain the equation for the critical temperature: 
\bea
C\left(\frac{T_c}{T_c(0)}\right)^{(2 - 2d)}=J_z\left|\frac{\Gamma(d/2 + i\alpha b_zH_y/T_c)}{\Gamma(1 - d/2 +i\alpha b_zH_y/T_c)}\right|^2 
&+& J_y\left|\frac{\Gamma(d/2 + i\alpha b_yH_z/T_c)}{\Gamma(1 - d/2 +i\alpha b_yH_z/T_c)}\right|^2\nonumber\\
C = (J_z + J_y)\left|\frac{\Gamma(d/2)}{\Gamma(1 - d/2)}\right|^2   
, &  & \alpha  = ev/2\pi c \label{critfield}
\eea

\begin{figure}
\begin{center}
\epsfig{file=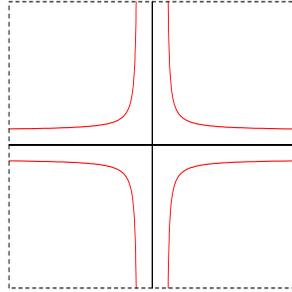,totalheight=1.5in,width=1.5in};
\caption{Angular dependence of the critical magnetic field.  This is plotted
for $d=1/2$.  The graph is qualitatively similar for other values of $d$.}
\label{hctheta}
\end{center}
\end{figure}

 The solution of this equation describes several interesting effects. 
\begin{itemize}
\item 
 A possibility of a re-entrance behaviour. 

 Let us consider the case when in-plane interactions are isotropic: $J_z = J_y, b_z = b_y$ and the magnetic field is directed at 45$^o$ angle $H_z = H_y = H$. This gives it the maximal power to suppress $T_c$. 
A numerical solution of Eq.(\ref{critfield})  is plotted in figure \ref{tch} (a) for
various values of the scaling dimension  $d$.  We see that there is a range of magnetic
fields for which the superconductivity exists in an intermediate range of temperatures.  To study the stability of these solutions one needs to have a good description of the ordered state in magnetic field, which we hope to obtain in the future. 

 At $T_c \rightarrow 0$ Eq.(\ref{critfield}) can be solved analytically which allows us to extract the value of critical field at $T_c = 0$: 
\bea
H_c(0) = \frac{2\pi c}{e}\frac{T_c(0)}{ bv}
\left(\frac{\Gamma (1-d/2)}{\Gamma (d/2)}\right)^{1/(1-d)}
\eea
This is plotted in figure \ref{tch} (b) along with the numerical solution
 for H$_c^{max}$.

\begin{figure}
\begin{center}
\epsfig{file=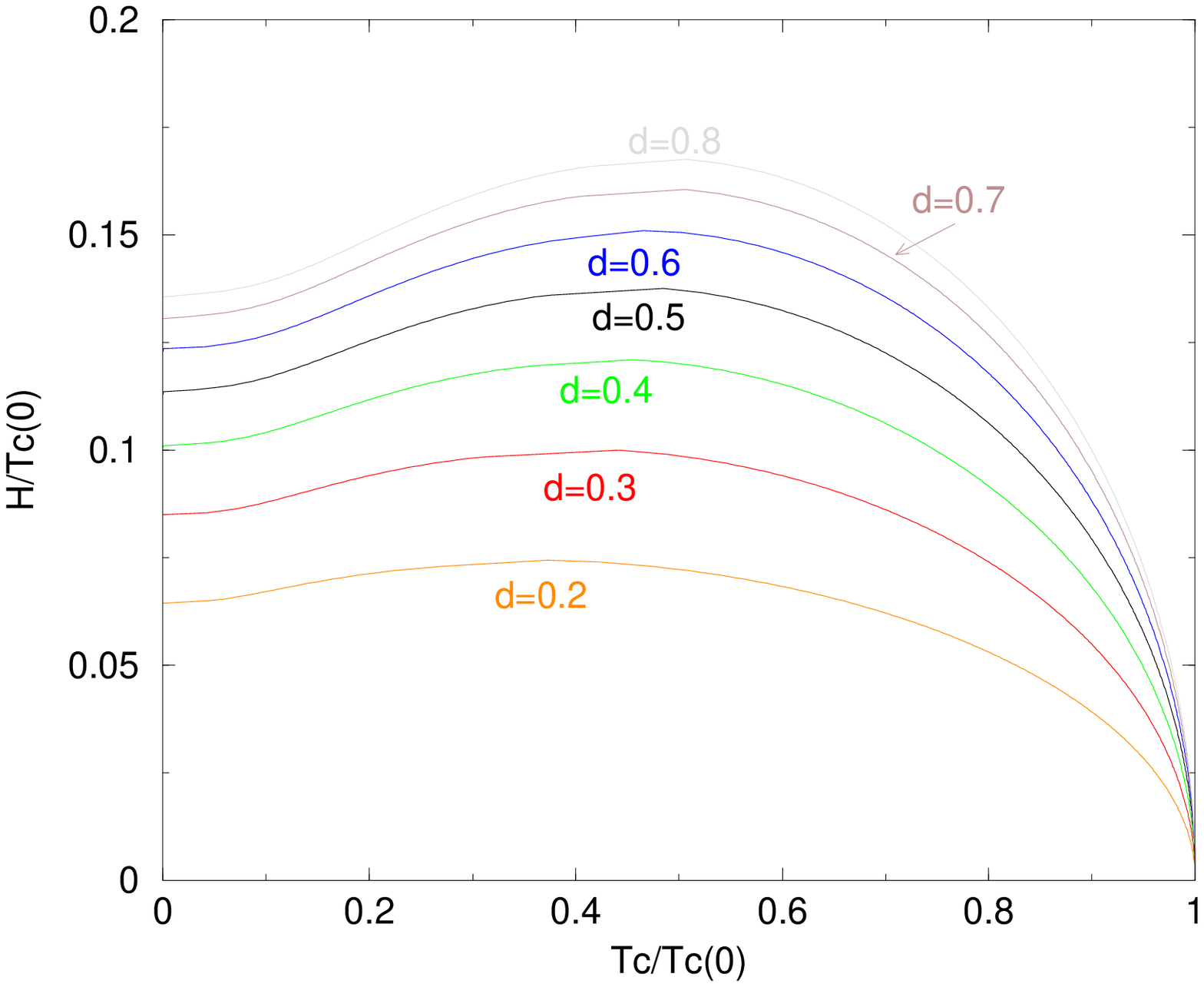,totalheight=3.0in,width=3.0in} \hspace{.5in}
\epsfig{file=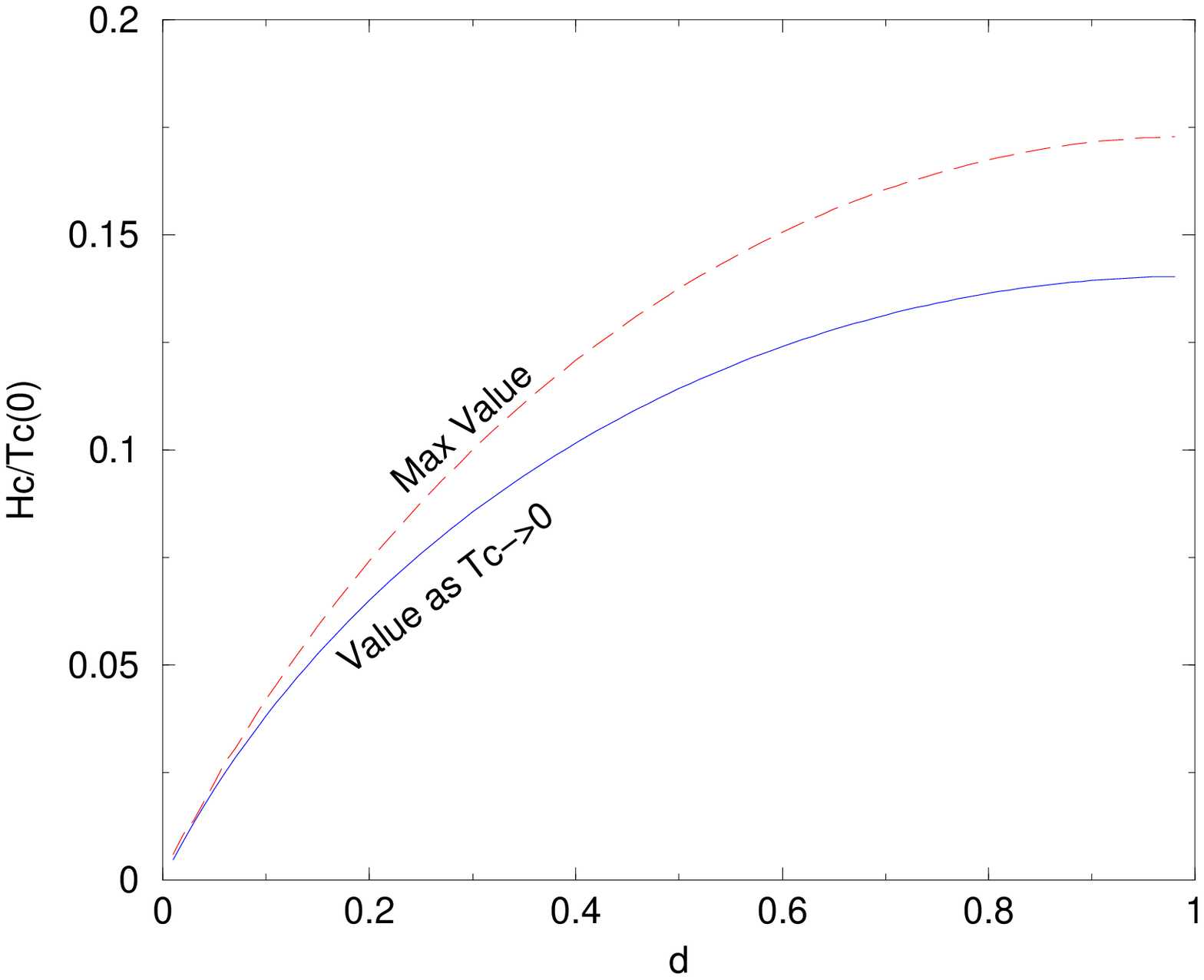,totalheight=3.0in,width=3.0in}
\end{center}
\caption{(a) The critical temperature as a function of magnetic field for various values of $d$.  (b) The critical magnetic field as a function of $d$. The magnetic field is measured in the units of $2e\bar h bv/c$. }
\label{tch}
\end{figure}

\item
Anisotropy of the phase diagram.
 
 Another prediction following from Eq.(\ref{critfield}) is an anisotropy of the phase diagram. This can be illustrated by an analytical solution for  $T_c \rightarrow 0$ case. Setting  $T_c \rightarrow 0$ Eq.(\ref{critfield}) in we find 
\bea
\frac{J_z}{(\alpha H_yb_z)^{2(1 -d)}} + \frac{J_y}{(\alpha H_zb_y)^{2(1 -d)}} = \frac{C}{[T_c(0)]^{2(1 -d)}}
\eea
This is plotted in figure \ref{hctheta}.  We must be careful to remember
however that this is a first order mean field calculation, and further
corrections will give a critical flux in all directions, even when the field
is pointing directly along one of the crystal axis.

\item SC-CDW transition.

 The validity of the above calculations is limited by the range of temperatures where the system is stable against CDW transition. Therefore, strictly speaking, before the
$T_c \rightarrow 0$ quantum critical point is reached the system will undergo a transition into a CDW state. 

\end{itemize}

\section{Corrections to RPA}\label{RPA2}

 The analysis of the previous sections was based on RPA. Since in realistic situations the number of nearest neighbours is never large, it is important to check how robust RPA is. We will calculate corrections to RPA in the simplest case  case of zero magnetic field. We shall also restrict ourselves to 
 $K_c=1$\ ($d = 1/2$ for both interactions).

\begin{figure}
\begin{center}
\psfig{file=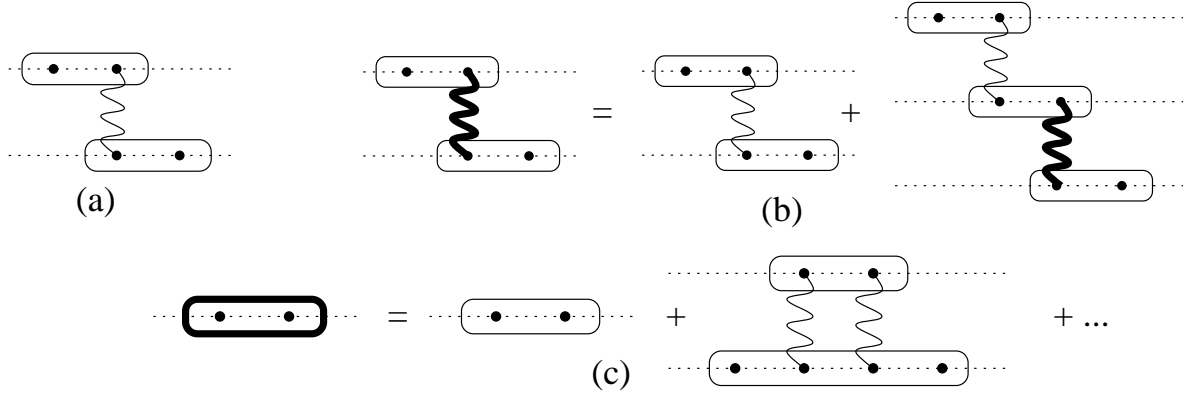}
\vspace{.1in}
\caption{(a) The basic RPA diagram, (b) The Dyson series for RPA, 
(c) The first correction term.  In
these diagrams, the dashed lines represent the 1D chains, the dots indicate
vertex operators of $\phi$ or $\theta$ and the wiggly lines are the inter-chain
interactions, and each diagram is an irreducible correlator.}
\label{rpa}
\end{center}
\end{figure}

The basic RPA calculation involves only the stronger of the
two interactions - for clarity let us again take this to be J.
  However as we mentioned before we would
 expect the presence of
the other competing interaction of the same scaling dimension to also play
 a role.  In particular we expect there to be a mode with a gap of $J-V$, seen
in (\ref{glvj}) and in the two chain model.
This will be very important around the point $V=J$ as it will become
massless thereby increasing fluctuations and decreasing the transition
temperature.
This can be investigated by looking at the first correction
 to the RPA formula - figure \ref{rpa}(c).

In terms of the fields $\phi$ and $\theta$, this diagram can be expressed as
\bea
\delta\chi & = & V^2 z_\perp\left[
\langle e^{i\sqrt{2\pi}\phi(a)}e^{i\sqrt{2\pi}\theta(1)}
e^{-i\sqrt{2\pi}\theta(2)}e^{-i\sqrt{2\pi}\phi(b)} \rangle
- \langle e^{i\sqrt{2\pi}\phi(a)}e^{-i\sqrt{2\pi}\phi(b)} \rangle
 \langle e^{i\sqrt{2\pi}\theta(1)}e^{-i\sqrt{2\pi}\theta(2)} \rangle \right]\nn
& & \hspace{3in} \times
  \langle e^{-i\sqrt{2\pi}\theta(1)}e^{i\sqrt{2\pi}\theta(2)} \rangle \nn
& + & J^2 z_\perp \left[
\langle e^{i\sqrt{2\pi}\phi(a)}e^{i\sqrt{2\pi}\phi(1)}
e^{-i\sqrt{2\pi}\phi(2)}e^{-i\sqrt{2\pi}\phi(b)} \rangle
- \langle e^{i\sqrt{2\pi}\phi(a)}e^{-i\sqrt{2\pi}\phi(b)} \rangle
 \langle e^{i\sqrt{2\pi}\phi(1)}e^{-i\sqrt{2\pi}\phi(2)} \rangle \right]\nn
& & \hspace{3in} \times
 ~ \langle e^{-i\sqrt{2\pi}\phi(1)}e^{i\sqrt{2\pi}\phi(2)} \rangle.
\eea

The revised RPA equation for the transition temperature is
\be
1 = \frac{Jz_\perp}{T_c} \left[ A_0^J + A_1^J \frac{J^2 z_\perp}{T_c^2}
+A_1^V \frac{V^2 z_\perp}{T_c^2} \right]
\ee
where the coefficients are given by
\bea
A_0^J & = & \frac{1}{\pi} \int_0^\pi d\tau \int_{-\infty}^\infty dx
\frac{1}{|\sinh (x+i\tau) |} = \frac{1}{2\pi}B^2(1/4,1/2) \\
A_1^J & = & \frac{1}{\pi^3} \int_0^\pi d\tau_1 d\tau_2 d\tau_b
\int_{-\infty}^\infty dx_1 dx_2 dx_b
\frac{1}{|\sinh(x_b+i\tau_b)|}\frac{1}{|\sinh(x_{12}+i\tau_{12})|^2} \nn
& & \hspace{2in}\times
\left[\frac{|\sinh(x_1+i\tau_1)||\sinh(x_{b2}+i\tau_{b2})|}
{|\sinh(x_2+i\tau_2)||\sinh(x_{b1}+i\tau_{b1})|} -1\right] \\
A_1^V & = &  \frac{1}{\pi^3} \int_0^\pi d\tau_1 d\tau_2 d\tau_b
\int_{-\infty}^\infty dx_1 dx_2 dx_b
\frac{1}{|\sinh(x_b+i\tau_b)|}\frac{1}{|\sinh(x_{12}+i\tau_{12})|^2} \nn
& & \hspace{1.5in} \times
\left[\left(
\frac{\sinh(x_1+i\tau_1)\sinh(x_2-i\tau_2)\sinh(x_{b2}+i\tau_{b2})
\sinh(x_{b1}-i\tau_{b1})}
{\sinh(x_1-i\tau_1)\sinh(x_2+i\tau_2)\sinh(x_{b2}-i\tau_{b2})
\sinh(x_{b1}+i\tau_{b1})}\right)^{1/2} -1\right]
\eea
with $x_{12}=x_2-x_1$ and so on.

The integrals are evaluated numerically be Monte-Carlo techniques,
\cite{num} with
values calculated over finite volumes then scaled to infinity.  The
results are
\bea
A_0^J & = & 4.377, \nn
A_1^J & = & 34.81 \pm 0.02, \nn
A_1^V & = & -33.01 \pm 0.02.
\eea

\begin{figure}
\begin{center}\hbox{
\psfig{file=tc3d3.ps,totalheight=3.0in,width=3.0in}\hspace{.5in}
\epsfig{file=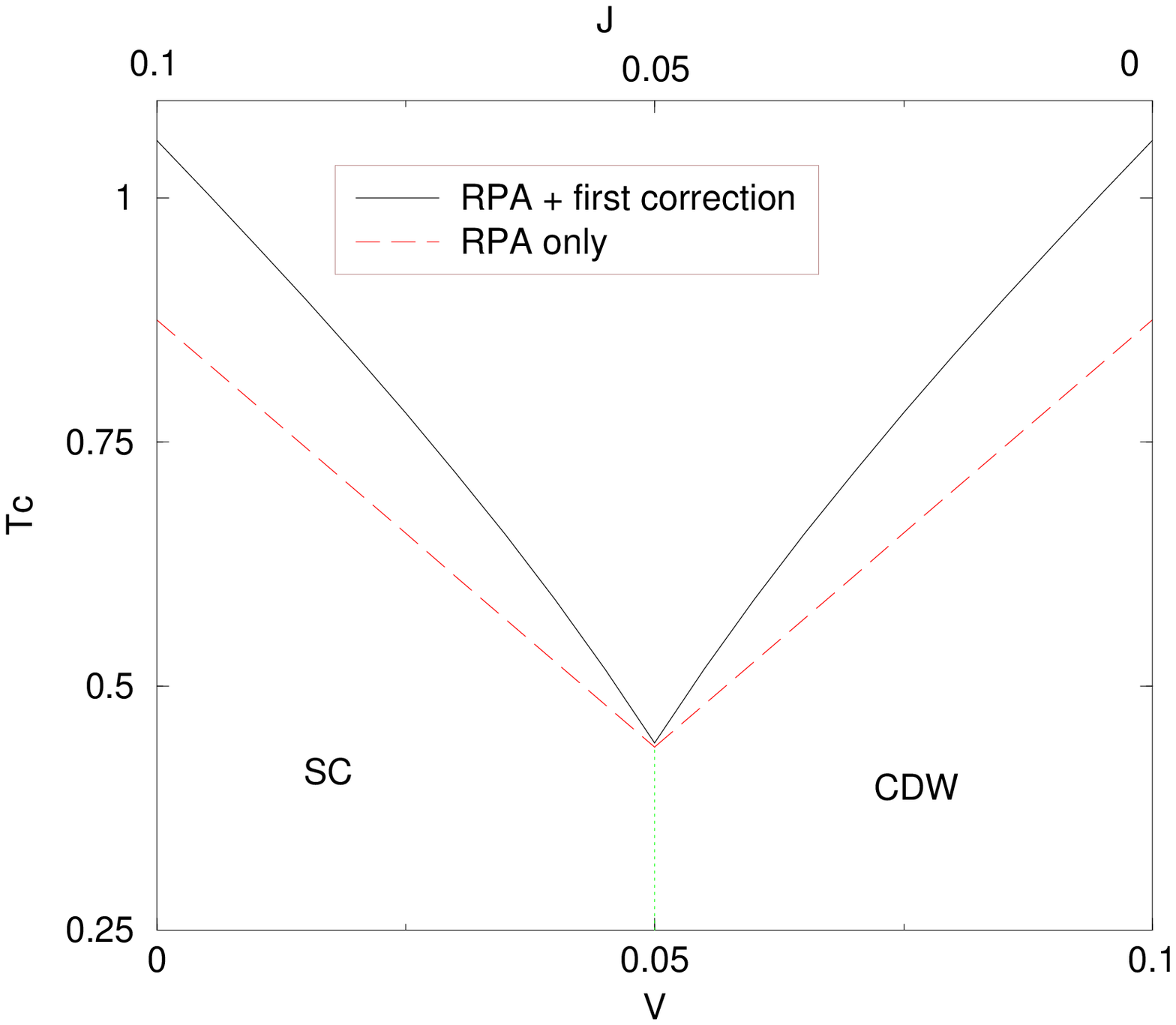,totalheight=3.2in,width=3.2in}}
\end{center}
\caption{(a) A plot of $T_c$ against $V$ and $J$,  (b) A cross section of
 $T_c$ against $V$ along the line $V+J=0.1$.  In these plots, we have taken
$z_\perp=2$ to allow these corrections to be clearly seen, although for
this approach to be valid, we require $z_\perp \geq 3$.}\label{tcvj}
\end{figure}

Hence the correction to the transition temperature is
\be
\frac{T_c}{A_0^J Jz_\perp} \approx 1 + \frac{1}{z_\perp}\left[ 0.42 - 0.40 \left(
\frac{V}{J}\right)^2\right].
\ee
This expression is valid for $J>V$.  If $V>J$, the expression is exactly
the same, but with $V\leftrightarrow J$.
This is plotted in figure \ref{tcvj} and gives a dip near the critical
point as expected.  

It is interesting to note that in the absence of the second interaction term,
ie $V=0$, these correction raise the transition temperature above the RPA
 value.  This differs from models of coupled spin
chains where RPA tends to overestimate the transition temperature.\cite{ik00,boc01}

\section{A Word about Two Dimensions}\label{rpa2d}

In two dimensions the RPA approach in the previous two sections must
break down completely, as spontaneous symmetry breaking is forbidden
by the Mermin-Wagner theorem.  We can see how this comes about by looking
at figure \ref{rpa}(c).  The correction we looked at involved
only bare couplings to the bare correlation function.  The process of
making these lines 'thick' involves much numerical complication and gives
rise to only small corrections in three or higher dimensions.
\cite{ik00}  However in
two dimensions, these corrections have infra-red divergences and drive
the transition temperature back down to $0$.

Nevertheless we still get a transition in two dimensions: it is of the 
Kosterlitz-Thouless\cite{ber70,kt73} type.  Let's look closer at Coulomb
 coupling in
two dimensions.  The Lagrangian for the coupled chains can be written
\be
{\cal L} = \sum_i \left\{ \frac{1}{2}(\partial_\mu \phi_i )^2
 - J\cos [ \beta (\phi_i - \phi_{i+1} ) ] \right\}.
\ee
By making the approximation 
\be
-\cos \phi = \frac{\phi^2}{2}\langle \cos \phi \rangle
\ee
which comes from the diagrammatic expansion, we can write this as
\be
{\cal L} = \sum_i \left\{ \frac{1}{2}(\partial_\mu \phi_i )^2
 + \tilde{J} (\phi_i - \phi_{i+1})^2 \right\}
\ee
with the self-consistent relation
\bea
\tilde{J} &=& J\beta^2\langle\cos  \beta (\phi_i - \phi_{i+1} ) \rangle \nn
&=& J\beta^2 \exp\left\{
-\beta^2\ T\sum_n \int \frac{dq_\perp}{2\pi}\frac{dq_\|}{2\pi}
\frac{1-\cos q_\perp}{\omega_n^2 + q_\|^2 + 4\tilde{J}\sin^2 (q_\perp/2)}
\right\}.
\eea

At $T=0$ this relation becomes
\bea
\tilde{J} &=& J\beta^2\exp\left(-\frac{\beta^2}{2\pi}\ln\frac{\Delta_s}{\sqrt{2\tilde{J}}}\right) \nn
& = & J\beta^2 \left(\frac{\tilde{J}}{\Delta_s}\right)^d
\eea
where $d=\beta^2/4\pi$ as before.  As $T$ increases, the self-consistent
value of $\tilde{J}$ will decrease, but for an estimate of the behaviour
of the transition temperature this relation will suffice.

The Kosterlitz-Thouless transition temperature \cite{boz,kt73}
$T_{\rm KT} \sim \sqrt{\tilde{J}} $ hence we have
\be
T_{\rm KT} \sim \Delta_s \left( \frac{J}{\Delta_s} \right)^{\frac{1}{2-2d}}
\ee
giving the same order of magnitude as the ordering temperature in higher
dimensions (\ref{tcrpa}).

\begin{figure}
\begin{center}
\psfig{file=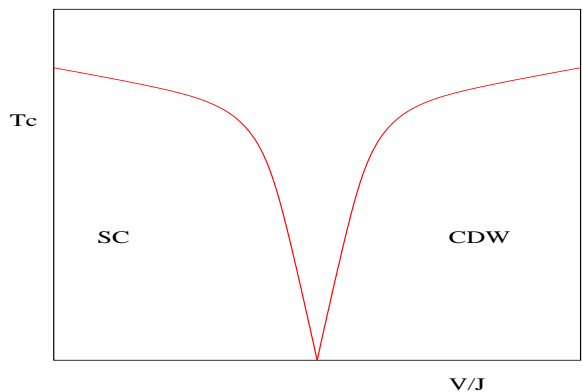, totalheight=2.0in, width=3.0in}
\caption{The modified phase diagram for our model in two dimensions}
\label{ktpd}
\end{center}
\end{figure}

Hence in two dimensions, although the nature of the transition is
different, the energy scales involved are the same as in higher dimensions.
  The only major difference occurs when approaching
the SU(2) critical point where the presence of a non-Abelian symmetry in two
 dimensions means that the transition temperature will drop to zero at this
point.  The qualitative phase diagram in two dimensions is shown in figure
\ref{ktpd}.

\section{An Example Experimental Systems}
\label{expt}

The class of materials ${\rm Sr_{12-x} Ca_{x} Cu_{24} O_{41}}$ are built
up from alternating layers of weakly coupled ${\rm Cu O_2}$ chains
 and ${\rm Cu_2 O_3}$ two-leg ladders.  The material shows a spin gap in
both of these one-dimensional units, \cite{tm+98}
 making it a prime candidate for
application of our model.  Our theory is still valid even if the superconductivity originates from  the ladders. 

For $x\geq 11.5$, these materials show superconductivity under
pressure, \cite{u+96,nu+98} and NMR\cite{tm+98} studies also indicate possible charge
 ordering at low
temperature and ambient pressure.  Recent measurements of the
electrodynamic response\cite{gh+02} have confirmed the presence of
CDW in this class of compounds.  One of the most interesting
measurements however is the DC resistivity.  For
${\rm Sr_{2.5} Ca_{11.5} Cu_{24} O_{41}}$ these measurements \cite{nu+98}
show a number of features:
\begin{itemize}
\item Below about 4 GPa pressure, the temperature dependence of the 
resistivity perpendicular and parallel to the ladders is different.  This
indicates that different mechanisms are governing the transport in these
two directions, consistent with the spin-gap concept.  Above 4 GPa the
 temperature dependence of the resistivity anisotropy  becomes weak,
which indicates that single particle hopping between ladders is now possible,
i.e. the spin gap has vanished and we have a crossover to a conventional  two-dimensional metallic 
behaviour.  This is consistent with the pressure dependence of the spin gap
 observed  in recent NMR experiments.\cite{pj+01}
\item At sufficiently high temperatures, coherent inter-ladder
charge dynamics is also seen.  The temperature where this occurs is consistent
with the NMR determinations of the spin gap, so we may conclude that
the transport properties of this material are indeed governed by weakly
interacting one-dimensional spin-gapped units.
\end{itemize}

\begin{figure}
\begin{center}
\psfig{file=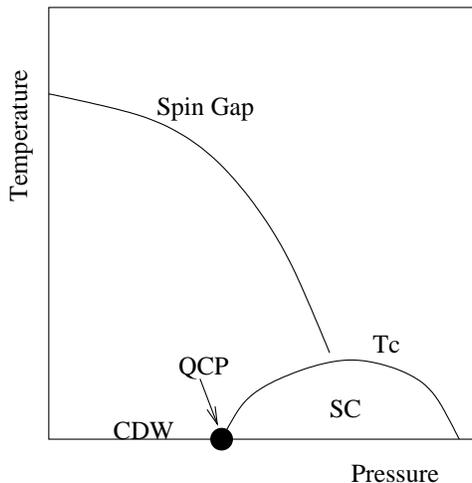, totalheight=2.5in, width=2.5in}
\caption{Qualitative phase diagram for the spin gap and superconducting
transition temperature against pressure in 
${\rm Sr_2 Ca_{12} Cu_{24} O_{41}}$}.
\label{pdtc}
\end{center}
\end{figure}

In figure \ref{pdtc} a qualitative phase diagram of this material is
shown. \cite{pj+01}  This is explained in terms of our model.
If we take $K_s \approx 1$ we have
\bea
 J_{\rm eff} & \sim & t^2/\Delta_s, \nn
 V_{\rm eff} & \sim & V_0 (\Delta_s/\Lambda).
\eea
The increase of spin  gap leads to decrease in the effective inter-ladder Josephson coupling. Hence eventually  the inter-ladder Coulomb interaction
 takes over and the 
superconductivity disappears. In quasi-two-dimensional the SC and CDW regions of the phase diagram are separated by  the quantum critical point, as described in Section VI. 

It would be interesting  for this material to measure the charge gap in the
superconducting region.  This may  be achieved via optical conductivity
measurements.  For the Luttinger liquid parameter $K_c \approx 1$, our model
then predicts the ratio $T_c/\Delta_c$ to be the non-BCS value of order of $0.4$.

Also in this material, $T_c$ is very small in comparison to the Fermi
 energy $v/a$, so the magnetic field effects on the superconducting state
should be strong.  This would be another interesting experiment to perform.

\section{Conclusion}

 We have discussed a model with the following hierarchy of energy scales: 
\begin{enumerate}
\item The highest energy scale is the spin gap $\Delta_s$. Below $\Delta_s$
 the system is described by competing CDW and SC fluctuations. 

\item There is a transition temperature at which either 
$\langle\cos\sqrt{2\pi}\Theta\rangle$ or $\langle\cos\sqrt{2\pi}\Phi\rangle$
 are formed. According to the mean field
 calculation, these order parameters cannot be formed simultaneously. Thus we
 are either in CDW or SC phase, but the temperature of their formation goes
 smoothly through the point $V = J$.

\item There is a third energy scale associated with the gap for another mode
 which becomes soft at the critical point.  This mode is not seen in the
first order RPA calculations, but it's effects can be noted by looking
at the first correction to RPA.
\end{enumerate}

Within the RPA approximation we calculated the transition temperature for
general $K_c$.  We calculated the ratio $T_c/M$ where $M$ is the zero
temperature gap in the charge sector.  We saw that this decreases below the
BCS value as the coupling strength is increased.  We also looked at the
properties of our model in a magnetic field, noting in particular the extreme
anisotropy of the phase diagram.

We then went on to calculate the first corrections to $T_c$
in the vicinity of the critical point which is decreased because of the
interplay between the two interactions.   We also showed that in two dimensions
where RPA breaks down completely, we get a transition of the
 Kosterlitz-Thouless type which has the same
energy scales as the ordering transition in higher dimensions.  We also showed
that the compound ${\rm Sr_2 Ca_{12} Cu_{24} O_{41}}$ is likely to be
described by our model and on this basis made further predictions about
it's properties and suggested that optical conductivity experiments should
be done on such a material.

S.T.C. acknowledges many useful discussions with Ralph Werner.  A.M.T. is
grateful to Marc Bocquet for giving him his preprint before publication.
The work is supported by the US DOE under
contract number DE-AC02-98 CH 10886 and EPSRC grant number 99307266.

\appendix

\section{Estimate of the Effective Couplings}
\label{vjest}

The easiest case is that of the Coulomb coupling.  In the bare system,
we have a term
\be
{\cal H}_{\rm coulomb} = \frac{V_0}{a_0}\sum_{n\ne m} \rho_n(x)\rho_m(x)
\ee
with $\rho(x)$ the charge density on each chain, and $V_0$ is the strength
of the inter-chain Coulomb coupling.
When we open a spin gap, two things happen to this expression.  Firstly,
anything involving the spin field is replaced by it's average value
$\langle\cos(\sqrt{2\pi}\Phi_s)\rangle \sim (\Delta_s/a_0^{-1})^{K_s/2}$.
Secondly, the cut-off in the normal ordering of the charge sector is
 changed from $a_0^{-1}$ to
$\Delta_s$.  This gives an extra factor of $(\Delta_s/a_0^{-1})^{K_c/2}$ for
each operator.
Overall, we generate an effective interaction
\bea
{\cal H}_{\rm cdw} = \frac{1}{2}\frac{V_{\rm eff}}{\Delta_s^{-1}}\sum_{n \neq m}
 :\cos[\sqrt{2\pi}(\Phi_n - \Phi_m)]: 
\eea
where
\be
V_{\rm eff} \sim \left(\frac{\Delta_s}{a_0^{-1}}\right)^{K_s+K_c -1} V_0.
\ee
In this paper, we will be keeping $K_s \approx 1$.

In the case of the effective Josephson coupling, we start from a single
particle
hopping term in our bare Hamiltonian density
\be
{\cal H}_{\rm hopping} = \frac{t}{2a_0} \sum_{n\ne m}
\left\{ R^\dagger_n R_m + L^\dagger_n L_m \right\}.
\ee
After opening the spin-gap, the effective Hamiltonian density only
involves pair hopping:
\bea
{\cal H}_{\rm sc} = \frac{1}{2\Delta_s^{-1}}J_{\rm eff}\sum_{n \neq m}
 :\cos[\sqrt{2\pi}(\Theta_n - \Theta_m)]:.
\eea
These are virtual processes involving an intermediate energy $\Delta_s$, hence
the $J_{\rm eff}$ will have a factor $t^2/\Delta_s$.  We must also remember to
change the cut-off in the normal ordering, so the overall expression is
\be
J_{\rm eff} \sim \left(\frac{\Delta_s}{a_0^{-1}}\right)^{1/K_c-1}
\frac{t^2}{\Delta_s}
\ee

\section{Mean Field solution for Many Chains}
\label{mcmf}

In the mean field approximation, the interaction term is
\bea
{\cal L}_{\rm int}&=&\sum_{m}\{V\cos[\sqrt{2\pi}(\Phi_n - \Phi_m)] +
 J\cos[\sqrt{2\pi}(\Theta_n - \Theta_m)]\} \nn
&\approx& z_\perp V\langle\cos[\sqrt{2\pi}\Phi]\rangle\cos[\sqrt{2\pi}\Phi_n] +
 z_\perp J\langle\sin[\sqrt{2\pi}\Theta]\rangle\sin[\sqrt{2\pi}\Theta_n]. 
\eea
 This can be written as
\bea
{\cal L}_{\rm int}&=&\sqrt{A^2 + B^2}\mbox{Tr}[(\cos\gamma I +
 \ri \sigma^1\sin\gamma)g + c.c], \nn
 A& =& Vz_\perp \langle \cos[\sqrt{2\pi}\Phi] \rangle, ~~~
 B = Jz_\perp\langle\sin[\sqrt{2\pi}\Theta]\rangle,~~~
 \tan\gamma = \frac{B}{A}
\eea
The constant matrix can be removed by the redefinition of $g$. After that it
 becomes evident that the free energy depends only on $R^2 = A^2 + B^2$.
The mean field equations are 
\bea
 A& =& - Vz_\perp \frac{\p F}{\p A} = -Vz_\perp\frac{A}{R}\frac{\p F}{\p R}, \nn
 B& =& - Jz_\perp \frac{\p F}{\p B} = - Jz_\perp\frac{B}{R}\frac{\p F}{\p R}   
\eea
From this it is clear that the only case where both $A$ and $B$
 are simultaneously non-zero is $V = J$. 

\bibliographystyle{../prsty}
\bibliography{../mybib.bib}

\begin{thebibliography}{10}

\bibitem{boz}
A.O.Gogolin, A.A.Nersesyan, and A.M.Tsvelik, {\em Bosonization and Strongly
  Correlated Systems} (Cambridge University Press, Cambridge, UK, 1998).

\bibitem{eme79}
V.J.Emery,  in {\em Highly Conducting One-Dimensional Solids}, edited by R.
  J.T.Devreese and V. Doren (Plenum, New York, 1979), p.\ 327.

\bibitem{fb81}
A.M.Finkel'stein and S.A.Brazovsky, J. Phys. C {\bf 14},  847  (1981).

\bibitem{sch96}
H.Schulz, Phys. Rev. Lett. {\bf 77},  2790  (1996).

\bibitem{etd97}
F.H.L.Essler, A.M.Tsvelik, and G.Delfino, Phys. Rev. B {\bf 56},  11001
  (1997).

\bibitem{betg01}
M.Bocquet, F.H.L.Essler, A.M.Tsvelik, and A.O.Gogolin, Phys. Rev. B {\bf 64},
  94425  (2001).

\bibitem{el74}
K.B.Efetov and A.I.Larkin, Soviet Physics JETP {\bf 39},  1129  (1974).

\bibitem{el75}
K.B.Efetov and A.I.Larkin, Soviet Physics JETP {\bf 42},  390  (1975).

\bibitem{ik00}
V.Y.Irkhin and A.A.Katanin, Phys. Rev. B {\bf 61},  6757  (2000).

\bibitem{boc01}
M.Bocquet, cond-mat/0110429  (2001).

\bibitem{coke00}
E.W.Carlson, D.Orgad, S.A.Kivelson, and V.J.Emery, Phys. Rev. B {\bf 62},  3422
   (2000).

\bibitem{ekt99}
V.J.Emery, S.A.Kivelson, and J.M.Tranquada, Proc. Natl. Acad. Sci. {\bf 96},
  8814  (1999).

\bibitem{ekz99}
V.J.Emery, S.A.Kivelson, and O.Zachar, Phys. Rev. B {\bf 59},  15641  (1999).

\bibitem{ek95}
V.J.Emery and S.A.Kivelson, Nature {\bf 397},  410  (1995).

\bibitem{wit84}
E.Witten, Commun. Math. Phys. {\bf 92},  455  (1984).

\bibitem{id89}
C.Itzykson and J-M.Drouffe, {\em Statistical Field Theory} (Cambridge
  University Press, Cambridge, UK, 1989), Vol.~2.

\bibitem{snt95}
D.G.Shelton, A.A.Nersesyan, and A.M.Tsvelik, Phys. Rev. B {\bf 53},  8521
  (1996).

\bibitem{sb83}
H.J.Schulz and C.Bourbonnais, Phys. Rev. B {\bf 27},  5856  (1983).

\bibitem{lz96}
S.Lukyanov and A.B.Zamolodchikov, Nucl. Phys. B {\bf 493},  571  (1996).

\bibitem{num}
W.H.Press, S.A.Teukolsky, W.T.Vetterling, and B.P.Flannery, {\em Numerical
  Recipes in C} (Cambridge University Press, Cambridge, UK, 1992).

\bibitem{ber70}
V.L.Berezinskii, Soviet Physics JETP {\bf 32},  493  (1970).

\bibitem{kt73}
J.M.Kosterlitz and D.J.Thouless, J. Phys. C {\bf 6},  1181  (1973).

\bibitem{tm+98}
M.Takigawa, N.Motoyama, H.Eisaki, and S.Uchida, Phys. Rev. B {\bf 57},  1124
  (1998).

\bibitem{u+96}
M. Uehara, T. Nagata, J. Akimitsu, H. Takahashi, N. Mori, and K. Kinoshita,
  Phys. Soc. Jpn. {\bf 65},  2764  (1996).

\bibitem{nu+98}
T.Nagata, M.Uehara, J.Goto, J.Akimitsu, N.Motoyama, H.Eisaki, S.Uchida,
  H.Takahashi, T.Nakanishi, and N.Mori, Phys. Rev. Lett. {\bf 81},  1090
  (1998).

\bibitem{gh+02}
B.Gorshunov, P.Haas, M.Dressel, T.Vuletic, B.Hamzic, S.Tomic, J.Akimitsu, and
  T.Nagata, cond-mat/0201413  (2002).

\bibitem{pj+01}
Y.Piskunov, D.Jerome, P.Auban-Senzier, P.Wzietek, C.Bourbonnais, U.Ammerhal,
  G.Dhalenne, and A.Revcolevschi, cond-mat/0110559  (2001).

\end{thebibliography}

\end{document}